\documentclass[9pt,twocolumn]{extarticle}
\pdfoutput=1

\usepackage{soul}
\usepackage{titlesec} 

\usepackage[backend=bibtex,style=nature]{biblatex}
\DeclareNameAlias{sortname}{first-last} 
\DeclareNameAlias{default}{first-last}
\bibliography{bibliography} 
\usepackage{dsfont}
\usepackage{graphicx}
\usepackage{subcaption}
\usepackage[margin=0.9in]{geometry}   
\usepackage[usenames,dvipsnames]{color}
\usepackage[colorlinks,linkcolor=Blue,urlcolor=Blue,citecolor=Blue]{hyperref}
\usepackage{amsmath,amssymb,braket}
\usepackage{amsfonts}
\usepackage{bbm}
\usepackage[squaren]{SIunits}
\usepackage{pifont}

\usepackage{array}
\newcolumntype{P}[1]{>{\arraybackslash}p{#1}}
\newcolumntype{Q}[1]{>{\centering\arraybackslash}p{#1}}

\usepackage{diagbox}

\usepackage{mathpazo}

\usepackage{courier}
\normalfont

\usepackage[T1]{fontenc}
\usepackage{caption}  
\usepackage{upgreek}

\newcommand{\ad}[1]{\textsuperscript{#1}\kern-2pt}

\usepackage{amsmath,amsthm,mathtools}

\usepackage{capt-of}

 \usepackage{flushend}

\usepackage[ruled]{algorithm2e} 
\usepackage{newfloat,algcompatible} 
\newcommand{\bracket}[3]{\langle#1|#2|#3\rangle}

\makeatletter
\newcommand\org@hypertarget{}
\let\org@hypertarget\hypertarget
\renewcommand\hypertarget[2]{%
  \Hy@raisedlink{\org@hypertarget{#1}{}}#2%
  }
\makeatother

\def\mytitle{Observation of Genuine High-dimensional Multi-partite Non-locality in Entangled Photon States
\vspace{-2mm}}      
 
\title{\vspace{-1.cm}\huge\textbf{\textrm{\mytitle}}}

\author{Xiao-Min Hu$^{1,2,3,\star}$, Cen-Xiao Huang$^{1,2,3,\star}$, Nicola d’Alessandro$^{4}$, Gabriele Cobucci$^{4}$, Chao Zhang$^{1,2}$,\\ Yu Guo$^{1,2,3}$, Yun-Feng Huang$^{1,2,3}$, Chuan-Feng Li$^{1,2,3}$, Guang-Can Guo$^{1,2,3}$,\\ Xiaoqin Gao$^{5,6}$, Marcus Huber$^{7,8}$, Armin Tavakoli$^{4,\dagger}$, Bi-Heng Liu$^{1,2,3,\ddagger}$}

\date{} 
\begin{document}

\twocolumn[
\maketitle 
\vspace{-9mm}
\begin{center}
\begin{minipage}{1\textwidth}
\begin{center}
\textit{\textrm{
\textsuperscript{1} Laboratory of Quantum Information, University of Science and Technology of China, Hefei, 230026, China.
\\\textsuperscript{2} CAS Center For Excellence in Quantum Information and Quantum Physics, University of Science and Technology of China, Hefei, 230026, China.
\\\textsuperscript{3} Hefei National Laboratory, University of Science and Technology of China, Hefei 230088, China.
\\\textsuperscript{4} Department of Physics and NanoLund, Lund University, Box 118, 22100 Lund, Sweden.
\\\textsuperscript{5} National Key Laboratory of Solid State Microstructure, School of Physics, Nanjing University, Nanjing, Jiangsu, 210093, China.
\\\textsuperscript{6} Collaborative Innovation Center of Advanced Microstructures, Nanjing University, Nanjing, Jiangsu, 210093, China.
\\\textsuperscript{7} Vienna Center for Quantum Science and Technology,
Atominstitut, Technische Universit¨at Wien, 1020 Vienna, Austria.
\\\textsuperscript{8} Institute for Quantum Optics and Quantum Information, Austrian Academy of Sciences, 1090 Vienna, Austria.
\\\textsuperscript{$\star$} These authors contributed equally: Xiao-Min Hu, Cen-Xiao Huang.
\\ ~~~Emails to: \textsuperscript{$\dagger$} armin.tavakoli@teorfys.lu.se; \textsuperscript{$\ddagger$} bhliu@ustc.edu.cn.}}
\end{center}
\end{minipage}
\end{center}
%\textsuperscript{$\dagger$} 
\setlength\parindent{12pt}
\begin{quotation}
\noindent 
{\bf{Quantum information science has leaped forward with the exploration of high-dimensional quantum systems, offering greater potential than traditional qubits in quantum communication and quantum computing. To advance the field of high-dimensional quantum technology, a significant effort is underway to progressively enhance the entanglement dimension between two particles. An alternative effective strategy involves not only increasing the dimensionality but also expanding the number of particles that are entangled. We present an experimental study demonstrating multi-partite quantum non-locality beyond qubit constraints, thus moving into the realm of strongly entangled high-dimensional multi-particle quantum systems.
%Our experimental approach employs the polarization-path degree of freedom to prepare both three- and four-particle Greenberger–Horne–Zeilinger (GHZ) states in three-level systems.
In the experiment, quantum states were encoded in the path degree of freedom (DoF) and controlled via polarization, enabling efficient operations in a two-dimensional plane to prepare three- and four-particle Greenberger–Horne–Zeilinger (GHZ) states in three-level systems.
Our experimental results reveal ways in which high-dimensional systems can surpass qubits in terms of violating local-hidden-variable theories. Our realization of multiple complex and high-quality entanglement technologies is an important primary step for more complex quantum computing and communication protocols.
}} 
\end{quotation}]

\noindent
Two-level quantum systems, known as qubits, predominantly serve as the fundamental unit of quantum information processing. A single qubit already reveals fundamental quantum principles, such as superposition and limitations on copying (i.e. no-cloning) \supercite{wootters1982single}, which propel applications in quantum cryptography \supercite{RevModPhys.81.1301}. However, by increasing the number of levels in the system, i.e.~the dimension, and the number of particles in the system, a lot of new physics and applications emerge.

Quantum particles with dimensions greater than two reveal new physical phenomena and applications, such as quantum contextuality \supercite{RevModPhys.94.045007} and noise-resistance quantum key distribution \supercite{PhysRevLett.88.127902}. Similarly, systems involving two particles lead to fundamental concepts such as entanglement \supercite{RevModPhys.81.865} and the device-independent framework for quantum information processing\supercite{liu2018device}. 
Progressing to a system of three or more particles has led to stronger forms of entanglement\supercite{greenberger1989going}, which has catalyzed the advent of quantum teleportation \supercite{hu2023progress} and quantum computing \supercite{o2007optical}. 
Systems that feature both a higher dimension and many particles lead to a more complex landscape of entanglement\supercite{PhysRevLett.110.030501, cobucci2024genuinely}, generate new non-local models \supercite{Lawrence2017, PhysRevLett.120.180402} and implement efficient quantum information tasks~\supercite{reimer2019high, wang2020qudits}.

Most studies on multi-qubit systems focus on increasing the number of entangled particles among which photons, ion traps, and superconducting circuits have achieved genuine multi-particle entanglement between 12 \supercite{PhysRevLett.121.250505}, 32 \supercite{PhysRevX.13.041052}, and 51 \supercite{cao2023generation} qubits, respectively. With the breakthroughs of experimental technology, the non-locality of multi-particle qubit systems has been observed in different systems \supercite{Barreiro2013,  erven2014experimental,baumer2021demonstrating}. 

\begin{figure}[th!]
\includegraphics [width= 0.48\textwidth]{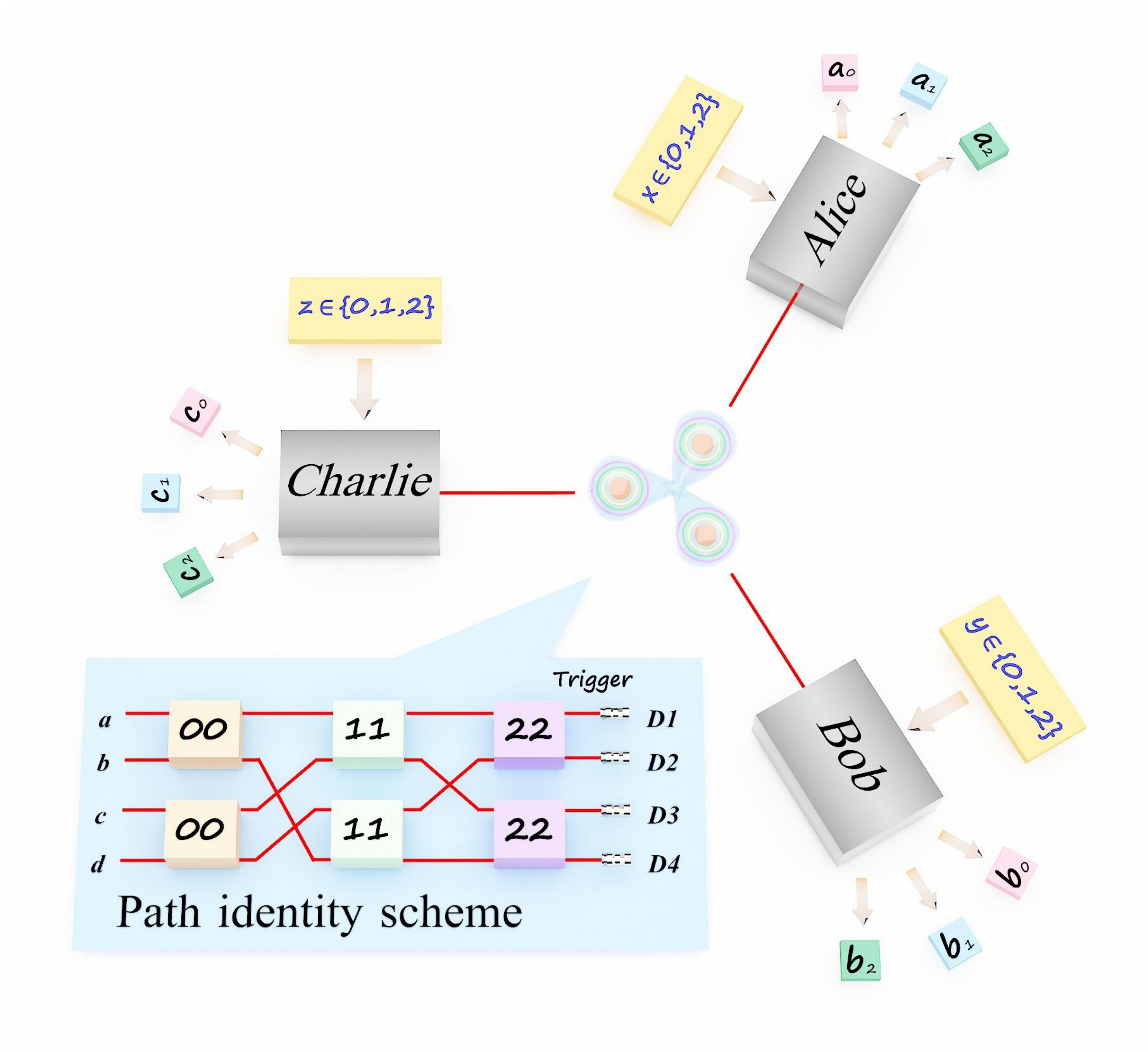}
\vspace{-0.3cm}%
\caption{\textbf{Three-dimensional three-particle GHZ state by Path Identity}. 
%(a) Three-dimensional maximally entangled state $|\phi\rangle=(|00\rangle+|11\rangle+|22\rangle)^{\otimes 2}/3$.  
Following the Path Identity scheme ~\supercite{krenn2017entanglement}, we use path exchange to achieve the preparation of high-dimensional GHZ states.
Each box means a crystal generates one pair of photons.
There are four photons in total, labeled by $a$ to $d$. The digital numbers $0$ to $2$ are the paths.
By exchanging the photons $b$, $c$, and $d$, the three-dimensional four-particle GHZ state can be generated if four detectors (D1-D4) click simultaneously. 
By projecting photon $a$ as a trigger onto the $(|0\rangle+|1\rangle+|2\rangle)/\sqrt{3}$ basis, we obtain the three-dimensional three-photon GHZ state. Then three photons are distributed to Alice, Bob, and Charlie, they randomly select the measurement basis (x, y, z) and obtain the results ($a_0,a_1,a_2...c_0,c_1,c_2$) to test the genuine high-dimensional multi-partite non-locality. }
\label{fig:scheme}
\vspace{0.2cm}%
\end{figure}

In recent years, there has been a growing recognition of the important role that high-dimensional quantum states play in probing the depths of fundamental physics and advancing quantum information applications \supercite{FriisNatPhys19,erhard2020advances}. High-dimensional quantum systems have a larger information capacity \supercite{PhysRevLett.76.4656,PhysRevLett.76.4656}, better noise resistance for quantum information tasks \supercite{PhysRevLett.127.110505, Ecker2019} and they simplify the complexity of quantum computing \supercite{lanyon2009simplifying, fedorov2012implementation, reimer2019high, quditcomputationringbauer}. Quantum non-locality in high-dimensional multi-particle systems is a natural endeavour. However, no high-dimensional multi-partite non-locality has been observed in experiments.

The difficulty is due to a combination of theoretical and technological challenges. In contrast to weaker forms of quantum correlations, genuine high-dimensional multi-partite non-locality is particularly challenging to characterize \supercite{PhysRevLett.115.020501, Tavakoli2019}. The challenge stems from the ambition to show that no quantum experiment based on multi-qubit states can account for the observed phenomena. Detection criteria must resolve this while simultaneously allowing for a significant level of noise-resistance. The latter is crucial, because high-dimensional multi-particle entanglement requires complex setups and it is difficult to prepare the most interesting states, such as GHZ states, at high fidelity. The challenge of preparing GHZ states of high quality is reflected in previous experiments where the fidelity was sufficient to witness entanglement but not to detect non-locality \supercite{erhard2018experimental,PhysRevApplied.17.024062,bao2023very}.

Here, as illustrated in Fig.~\ref{fig:scheme},
%we use the method of coupling the polarization and path degrees of freedom in photons to 
%\textcolor{red}{we experimentally demonstrate high-fidelity high-dimensional multi-particle entanglement.}
we leverage the concept of path identity to encode quantum states into the path DoF. By utilizing polarization to control path exchanges, we experimentally achieve the preparation of high-fidelity, high-dimensional multi-partite entangled photons (details see Section 8 in the SM).
We report on the observation of entanglement that is strong enough to exhibit genuine high-dimensional multi-particle non-locality through the explicit violation of Bell-type inequalities . 

%%%%%%%%%%%%%%%%%%%%%%%%%%%%%%%%%%%%%%%%%%%%%%%%%%%%%%%%%

\section*{Results}
Compared to multi-qubit states, the structure of high-dimensional multi-particle quantum states is more complex. Combining high-dimensionality with many particles can, for instance, be revealed by a notion of dimensionality vector \supercite{PhysRevLett.110.030501}. The smallest element of this vector can be interpreted as the genuine multi-partite entanglement dimension;  representing the highest necessary dimensionality cost if the state were to be generated using only bipartite entanglement sources \supercite{cobucci2024genuinely}. The quality of multi-particle entanglement can, in principle, be benchmarked through fidelity measurements \supercite{Fickler2014, cobucci2024genuinely, bavaresco2018measurements}. However, this faces the obstacle that implementations typically have low count rates, which makes it resource-intensive to collect a large amount of statistics on many different bases.  Therefore, one typically tries to use as few bases as possible, such as only two mutually unbiased bases for all parties \supercite{bavaresco2018measurements}. We employ a recent improvement of such witnesses, also using only two global product basis measurements to bound the fidelity \supercite{cobucci2024genuinely} and thus also the genuine multi-partite entanglement dimension.  Moreover, we propose Bell-type inequalities for witnessing high-dimensional non-locality for three qutrits. The violation of these inequalities excludes not only local hidden variables but also quantum non-locality based on arbitrary qubit systems and non-locality based on systems where some particles are qubits and others are qutrits. These tests pave the way for practically viable device-independent tests of entanglement dimensionality in multipartite systems.

\subsection*{Entanglement detection via efficient fidelity estimation}

The standard method for verifying that a multi-particle state is fully entangled is to prove that it exhibits genuine multi-partite entanglement (GME); see e.g.~\supercite{Bourennane2004, Lu2007, Yao2012, Wang2016, PhysRevLett.121.250505, Thomas2022,FriisNatPhys19}. A state is called GME if it cannot be recreated by classically mixing states that are separable with respect to some bipartition of the set of particles.  However, when the photons have more than two DoFs, it is important to also verify that the state is genuinely high-dimensional, i.e.~that it does not permit a representation using only lower-dimensional (qubit) states. Formally, this means that Schmidt rank across all partitions is larger than two, for every possible decomposition of the density matrix \supercite{PhysRevLett.110.030501, FriisNatPhys19}. While a daunting task to prove in general, a simple criterion for an $n$-partite state, $\rho$, being both genuinely three-dimensional and GME is that its fidelity with an ideal $n$-photon three-dimensional GHZ state $|\text{GHZ}_{n,3}\rangle:=\frac{1}{\sqrt{3}}(\sum_i|i\rangle^{\otimes n})$ exceeds the following limitation \supercite{hu2020experimental,cobucci2024genuinely},
\begin{equation}\label{Fidelity_criterion}
F_{\text{GHZ}} = \bracket{\text{GHZ}_{n,3}}{\rho}{\text{GHZ}_{n,3}}\leq \frac{2}{3}.
\end{equation} As later discussed, we can efficiently test this criterion in our experiments using a three-photon GHZ state. 

However, measuring the fidelity becomes significantly more demanding for a larger number of high-dimensional photons due to an increased number of measurements. To overcome this, one can use a weaker but less resource-costly criterion, which uses only two complementary basis measurements. Such criteria are tailored for high-quality sources in the sense that they perform well if the experiment can generate a state close to the perfect GHZ state.  Specifically, let $\mathcal{C}=\{\ket{i}\}$ be the computational basis and $\mathcal{F}=\{\ket{f_i}\}$ be a complementary basis, obtained by a Fourier transform of the computational basis. Consider now that all $n$ photons are measured in the former basis and the measurement is considered successful if all outcomes are identical. The probability of success is called $p(\text{identical}|\mathcal{C})$. 
Next, we measure all photons in the complementary basis and consider it successful if the outcomes sum to zero modulo the dimension, three.
We denote that probability $p(\text{sum-zero}|\mathcal{F})$. Ref.~\supercite{cobucci2024genuinely} shows that a state is three-dimensionally GME if it violates the condition
\begin{equation}\label{GMEwitness}
    W=p(\text{identical}|\mathcal{C})+p(\text{sum-zero}|\mathcal{F})\leq \frac{5}{3},
\end{equation}
and that it implies the fidelity bound $F_\text{GHZ}\geq W-1$. 
One can verify that a perfect three-dimensional GHZ state achieves the violation $W=2$, thus corresponding to a perfect fidelity. For example, assume that the state generated in the experiment is a mixture of the GHZ state and isotropic noise. In our systems, we have $n=4$ particles of dimension $d=3$. A successful detection of three-dimensional GME  requires a GHZ state visibility of at least $79.5\%$.

\begin{figure*}[th!]
\includegraphics [width= 1\textwidth]{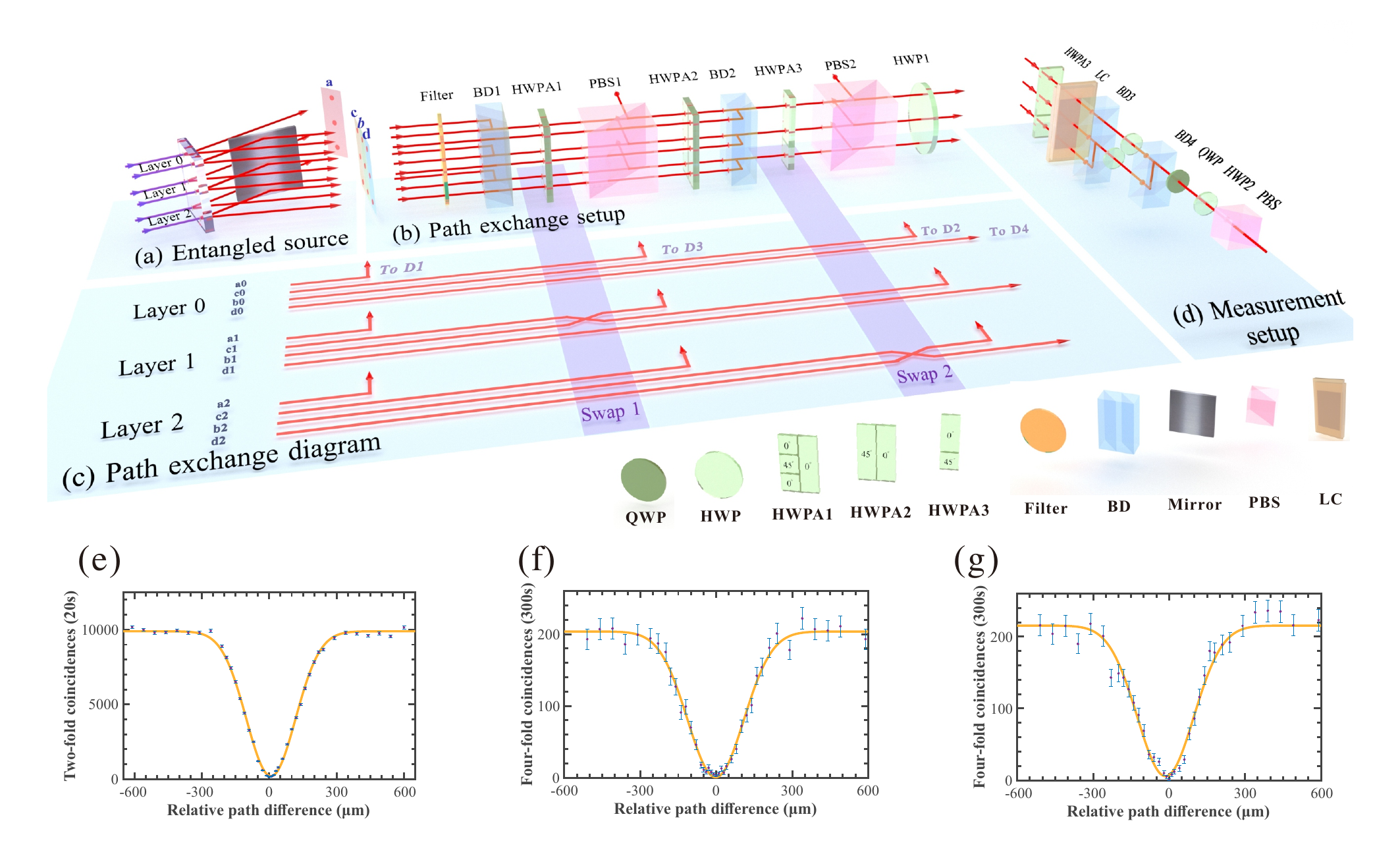}
\vspace{-0.3cm}%
\caption{\textbf{Schematic diagram of high-dimensional multi-photon GHZ state generation.} (a) Schematic diagram of entanglement source. The ultrafast pulsed laser (@775~nm) with a repetition rate of 76 MHz is evenly split into 6 beams ($3\times 2$ arrays) to pump the type-II barium borate (BBO) crystal, and generated 2 pairs of 3-dimensional entangled photons  $((|00\rangle+|11\rangle+|22\rangle)/\sqrt{3})^{\otimes 2}$, labeled by $a, b, c$ and $d$ respectively. We use a $5~cm \times 5~cm$ mirror to reflect the photon $c$, which is parallel to photons $b$ and $d.$ These photons are set equidistant by 4~mm apart.  (b) Experimental setup of Photon exchange. By changing the angles of HWPAs (1-3), we can achieve the exchange of paths $b$ and $c$ in layer 1 and paths $b$ and $d$ in layer 2. The PBSs (1-2) are used to separate four photons into four different paths. Therefore, the three-dimensional four-photon GHZ state is created. It is worth noting that in order to ensure that the photons participating in path exchange have better indistinguishability, we used a 3~nm filter at the beginning.
(c) Path exchange scheme. By exchanging path $|b1\rangle$ and path $|c1\rangle$ in layer 1, and path $|b2\rangle$ and path $|d2\rangle$ in layer 2, we can get the same structure as the path identity scheme in Fig. \ref{fig:scheme}. (d) Measurement setup. Any projective measurement can be performed by setting the HWPs, QWP, and a liquid crystal (LC) for each party. (e)-(g) Test of indistinguishability between different spontaneous parametric down-conversion (SPDC) photons. The Hong–Ou–Mandel (HOM) interference results of photon cd, bd and bc were shown in (e), (f) and (g), and observed HOM visibility of $0.981\pm 0.001$, $0.996\pm 0.004$ and $0.974\pm 0.008$, respectively.}
\label{fig:exchange}
\vspace{0.2cm}%
\end{figure*}

\subsection*{Multi-partite high-dimensional non-locality}
Consider a three-partite state distributed between Alice, Bob and Charlie. They independently select one of three possible measurements and perform them on their share of the state, each yielding one of three possible outcomes. For this three-party, three-input, three-output Bell scenario, Ref.~\supercite{Lawrence2017} developed a Bell inequality which can be maximally violated by basis measurements on a three-dimensional GHZ state. However, a Bell inequality violation does not necessarily imply that the non-locality must occur between three three-dimensional systems; the violation may be possible using already three-qubit entanglement. 

Using the techniques of \supercite{PhysRevLett.115.020501}, we have determined bounds for the Bell test parameter when the dimensionality of the first, second and third particle is, respectively, $(d_A,d_B,d_C)$. Our bounds take the form of a chain of inequalities,
\begin{equation}\label{nonlocalitybounds}
\mathcal{B} \stackrel{\text{LHV}}{\leq} 7\stackrel{\text{(2,2,2)}}{\leq} 7.446 \stackrel{\text{(2,2,3)}}{\leq} 7.584 \stackrel{\text{(2,3,3)}}{\leq} 8.225\stackrel{\text{(3,3,3)}}{\leq}9.
\end{equation}
Here, $\mathcal{B}$ is the Bell test parameter measured in the experiment. The first bound (LHV) is the standard Bell inequality from Ref.~\supercite{Lawrence2017}. We see that a significant portion of the quantum non-locality can be modeled by systems of only qubits. We also see that stronger violations can be attributed to systems of two qubits and one qutrit and one qubit and two qutrits, respectively. We remark that $\mathcal{B}$ is symmetric under party permutation, and therefore, the bounds apply for any permutation of the tuple $(d_A,d_B,d_C)$. A three-qutrit GHZ state achieves the maximal value, $\mathcal{B}=9$.  The bounds are tight up to several decimal points (not shown in Eq. \eqref{nonlocalitybounds}). Details on both the specific construction of $\mathcal{B}$ and the derivation of the bounds are given in Section 6 of the Supplementary Materials (SM). 

We aim to violate these Bell-type inequalities and thereby demonstrate genuine high-dimensional multi-particle non-locality. However, this requires high fidelity in the state generation. Considering again the noise model in which the experiments generate a mixture of a GHZ state and isotropic noise, the critical visibility of the GHZ state for achieving the chain of violations becomes $v_{\text{LHV}}=0.66$, $v_{222}=0.74$, $v_{223}=0.76$ and $ v_{233}=0.87$. We see that the task is particularly demanding for falsifying quantum non-local models based on one qubit and two qutrits. In our experiment, we report on violating all these inequalities using three-photon qutrit GHZ states.

\subsection*{Finite count analysis}
The probabilities entering into the tests of entanglement and nonlocality described in Eq.~\eqref{Fidelity_criterion}, \eqref{GMEwitness} and \eqref{nonlocalitybounds} cannot be precisely determined by experiments with finite count statistics. They can only be estimated as relative frequencies up to a confidence that depends on the number of times a measurement is performed, $N$. Since this is an important practical limitation, it is relevant to determine the confidence with which these tests can be violated for realistic values of $N$. This discussion is detailed in Section 7 in the SM. There, Chernoff's bound \cite{Chernoff,Dimi__2018} is used to bound the $N$ needed to ensure that at least a given confidence is obtained for an experimental violation of the inequality based on relative frequencies. Representing the confidence in terms of the $p$-value, this bound takes the form
\begin{equation}\label{Chernoff}
    N > \frac{1}{D_{KL}(\mathcal{W}_{B} + \Delta_{\text{exp}}||\mathcal{W}_{B})} \ln \frac{1}{p\text{-value}},
\end{equation}
where $\mathcal{W}_{B}$ is the inequality bound to violate for entanglement and non-locality detection, $\Delta_{\text{exp}}$ is the violation measured in the lab and $D_{KL}(\mathcal{W}_{B} + \Delta_{\text{exp}}||\mathcal{W}_{B})$ is the Kullback-Leibler divergence. The main feature here is that if the observed violation magnitude $\Delta_\text{exp}$ is large, then high confidence is already implied by reasonably small $N$. Such large violations are associated with high-quality entanglement and nonlocality certificates.

%\section*{Experimental three-dimensional four-photon GHZ Entanglement}
\subsection*{Experimental three-dimensional four-photon GHZ Entanglement}
The preparation of high-dimensional multi-photon entangled states has long been a challenge due to their complex structures and lack of a universal experimental preparation scheme. In our experiment, we used narrowband filtering to achieve high visibility interference between high-dimensional entangled photon pairs combined with a path exchange method of polarization-path coupling, to efficiently prepare a four-photon three-dimensional GHZ state using the path identity scheme \supercite{krenn2017entanglement}.
The path of photons can, in principle, encode infinite-dimensional quantum information. For a three-dimensional four-photon GHZ state, it can be written as:
\begin{equation}
|GHZ_{4,3}\rangle=\frac{1}{\sqrt{3}}(|0000\rangle+|1111\rangle+|2222\rangle)_{abcd},
\label{GHZ state}
\end{equation}
where $0$ to $2$ are the possible paths and $a$ to $d$ are the photons. As shown in Fig.~\ref{fig:scheme}, our experiment on a three-dimensional four-photon GHZ state was implemented with two steps: Photon pairs preparation and path exchange, shown in Fig. \ref{fig:exchange}(a) and Fig. \ref{fig:exchange}(b) respectively.  

The core idea of the path identity scheme is to achieve the preparation of high-dimensional multi-photon quantum states through indistinguishable photon path exchanging. Good indistinguishability of photons is a prerequisite for achieving high-fidelity quantum state preparation. 
In our experiment, a pulsed laser beam with an average power of $\sim 1.7$ W, a central wavelength of 775 nm, and a repetition rate of 76 MHz is split into six beams ($3\times 2$ array) to pump the type-II barium borate (BBO) crystals with a thickness of 6.3 mm, and the beams of each vertical layer are labeled as layers 0, 1, 2, which represent different spatial modes. Due to the difference in collection efficiencies, we balance the power of each pump beam, ensuring equal count rates for each source. Then, through spontaneous parametric down-conversion (SPDC), 2 pairs of 3-dimensional entangled photons  $((|00\rangle+|11\rangle+|22\rangle)/\sqrt{3})^{\otimes 2}$ are generated, labeled by $a, b, c$ and $d$ respectively (details see Section 2 in the SM). 
The experiment's high-performance light source ensures excellent photon indistinguishability between different paths. To ensure the indistinguishability of the four photons, we used a 3~nm filter to filter the three photons b, c, and d. The visibility of Hong–Ou–Mandel (HOM) interference between the three photons was $0.981\pm 0.001$ (cd), $0.996\pm 0.004$ (bd) and $0.974\pm 0.008$ (bc),  respectively. The excellent indistinguishability ensures that we have prepared high-dimensional multi-photon entangled states with high fidelity.

In our scheme, one can simply see the processes in Fig. \ref{fig:exchange} (c); we only need to do two path exchanges; path $b1$ and $c1$ are exchanged in layer 1, while path $b2$ and $d2$ are exchanged in layer 2.  Afterwards, if we choose to observe D1-4 coincidence, we can obtain four qutrit quantum states.

For exchanging photon paths, we use polarization devices to control the polarization and path DoFs of photons. This method is efficient, stable, and very suitable for the preparation of high-dimensional multi-photon entangled states that require long-term phase stability. As shown in Fig.\ref{fig:exchange}(a), we use a $5~cm \times 5~cm$ mirror to reflect photon $c$ such that it is parallel to photons $b$ ($d$) at a distance of 4~mm (8~mm). In Fig.\ref{fig:exchange}(b) path exchange setup, we first use a 3~nm narrow-band filter to increase the coherence length of the photons further. 

In our experiment, we achieve path exchange by continuously converting the polarization and path DoFs of photons. As shown in Fig.\ref{fig:exchange}(b), BD1 combines the photons in path c with photons in path b through polarization merging. The HWP array (HWPA1) is used to achieve second-layer photon polarization exchange in the b path. The photons in $H$ polarization in paths $b$ and $c$ pass through PBS1, while the photons in $V$ polarization reflect and reach $D3$. At this point, we have achieved the exchange of photons in paths $b$ and $c$ in layer 1. Then, BD1 is used to combine the photons in path d with photons in path b through polarization merging. The HWP3 is used to achieve layer 2 photon polarization exchange in the b path. Finally, we separated $H$-polarized photons and $V$-polarized photons through PBS2 and reached D2 and D4, respectively, achieving exchange on layer 2 along the b and d paths. If we label three layers as $\ket{0}$, $\ket{1}$, and $\ket{2}$ we then obtain a three-dimensional four-photon GHZ state \eqref{GHZ state}.
As shown in Fig.\ref{fig:exchange}(d), any projective measurement can be implemented by adjusting HWPs, QWPs and LCs in our experiment (details see Section 2 in the SM). 

At the same time, our experimental techniques can be easily extended to other graph states based on path identity schemes \supercite{PhysRevLett.119.240403, PhysRevA.99.032338} and further enhanced by leveraging the recently developed concept of spatial overlap among identical particles\supercite{PhysRevLett.120.240403,Lee:22,Barros:20}. This approach can be generalized to the linear graph method for efficient scalability\supercite{Chin2021graphpictureof,chin2024shortcut}. 
%Any projective measurement can be implemented by adjusting HWPs and two liquid crystals (LC1-2) in our experiment. LC loads a phase between $H$ polarization and $V$ polarization. This is a standard technique for measuring path quantum states; we  refer to the supplementary materials for details. 

%\section*{Experimental results}

%\textcolor{red}{In the experiment, we demonstrated the presence of GME in 3-qutrit and 4-qutrit case. Furthermore, we experimentally verified high-dimensional multipartite nonlocality for the first time. The detailed experimental conclusions are as follows:}

\subsection*{Three qutrit GME witness}
We use the previously described efficient fidelity witness \eqref{GMEwitness} to verify that our quantum state is genuinely high-dimensional and genuinely multi-photon entangled. We performed measurements of all photons in the computational basis ($\mathcal{C}$) and Fourier basis ($\mathcal{F}$), respectively. This represents a distinct advantage already for three qutrits; tomography would require $1728$ projection measurements whereas our witness requires only $3^3\times2=54$ projective measurements. Integrating for 200 seconds for each projection, a total of 1142 counts is recorded. We obtain the maximum violation of $W_{3qutrit}=1.849 \pm 0.016 > \frac{5}{3}$, which is sufficient to observe the genuine three-dimensional three-photon entanglement with a $p\text{-value}$ of order $10^{-18}$ according to \eqref{Chernoff}. The associated lower bound on the three-qutrit GHZ-state fidelity is $0.849\pm 0.016$. To confirm the accuracy of our fidelity estimation, we used subspace measurement methods to accurately estimate the three-particle qutrit state \supercite{erhard2018experimental}. To calculate $F_{exp}$, it is sufficient to measure the 32 diagonal and 6 unique real parts of off-diagonal elements of $\rho_{exp}$ (details see Section 1 in the SM). The four-fold coincidence rate here is about $\sim 1.1$ Hz, and the integration time of each measurement setting is 1000~s. From the experimental data, $F_{exp}$ is calculated to be  $0.910\pm 0.006$, which is above the bound of $F_{(3,3,2)} =0.667$ by 40.5 standard deviations. The high fidelity proves that the quantum states we prepare can be used for tasks such as quantum secret sharing \supercite{PhysRevA.59.1829} and distributed quantum sensing \supercite{liu2021distributed} in the future.

%We observed the GME of 3-qutrits and 4-qutrits in the experiment, respectively. 
%For the 3-qutrit case, we use 4 sets of measurement bases (calculation basis C, and 3 sets of MUBs $F_0 - F_2$) to construct the witness $W=2C+F^{000}+F^{012}+F^{102}$. For each measurement basis, we integrated for 200 seconds, and total record 2245 counts. We obtain the maximum violation of $W_{3 qutrit}=4.6125 \pm 0.0289 > 4$, which is sufficient to observe the genuine three-dimensional three-photon entanglement, and the lower bound of the (3,3,3) GHZ state fidelity is $0.8708\pm 0.0096$.
%For the case of 3-qutrits, there are $3^3\times2=54$ projective measurements \MH{for the simplified witness, compared to $1728$ measurements for tomography}. 

%\begin{figure}[tbph]
%\includegraphics [width=.5\textwidth]{Experimental_plot_fid_3qutrit.pdf}
%\vspace{-0.3cm}%
%\caption{\textbf{Experimental results for 3-qutrit GHZ fidelity.} Plot of the fidelity $F(\rho)$ between the experimentally produced state $\rho$ and the ideal 3-qutrit GHZ state. The dashed sections denote the bounds from criterion \eqref{Fidelity_criterion} for the detection of 3 GME-dimensions. The red dot is the experimental value.}
%\label{fig:plot_fid_3qutrit}
%\vspace{0.2cm}%
%\end{figure}

\subsection*{Four qutrit GME witness}

\begin{figure}[tb]
\includegraphics [width=.48\textwidth]{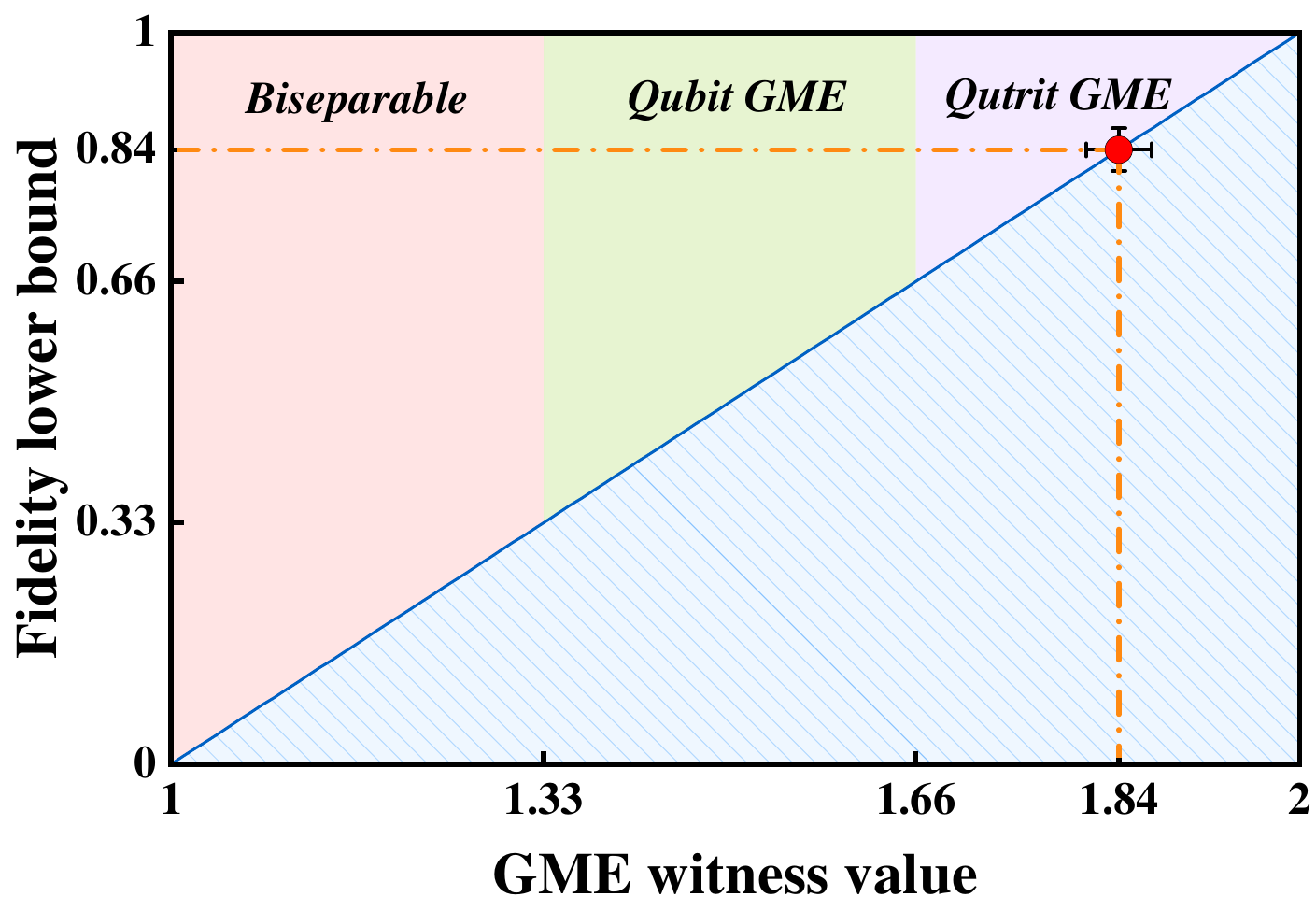}
\vspace{-0.3cm}%
\caption{\textbf{Experimental results for 4-qutrit GHZ witness.} The GHZ-state fidelity of the experimentally produced state is shown as a function of the GME witness value $W(\rho)$ given in \eqref{GMEwitness} for qutrit systems. The red,  green and purple regions correspond to biseparable states, two-dimensional GME and three-dimensional GME, respectively. 
The blue hatched region represents the fidelity lower bound $F_{\text{GHZ}} \geq W - 1$, which certifies the minimum fidelity achieved in our experiment. 
%The blue section refers to the fidelity lower bound $F_{\text{GHZ}} \geq W - 1$. \
The red dot is the experimental result, based on a total of  $N_{tot} = 2404$ counts. The statistical confidence in falsifying two-dimensional GME corresponds to a $p$-value on the order of $10^{-17}$ (details see Section 7 in the SM).}
\label{fig:plot_fid_4qutrit}
\vspace{0.2cm}%
\end{figure}

For the case of four qutrits, the efficient witness requires a total of $3^4\times2=162$ projective measurements, which can be compared to  $20736$ measurements for tomography, or even $324$ for complete fidelity measurements. For each measurement basis, we integrated for 1000 seconds, which led to a total of 2404 counts recorded for all measurements. We obtain the maximum violation of $W_{4 qutrit}=1.841\pm 0.029 > \frac{5}{3}$. Notably, for this violation, total counts $N \simeq 300$ would already be enough to get a $p$-value of order $10^{-5}$. Our total counts lead to a $p$-value $ \leq 10^{-17}$. The associated four-qutrit GHZ fidelity is then lower bounded by $0.841\pm 0.029$. The experimental result and its genuinely three-dimensional entanglement properties are illustrated in Figure~\ref{fig:plot_fid_4qutrit}.

The two above reported entanglement witness tests prove that we have successfully prepared genuine high-dimensional multi-photon entangled states in our experiment. 

%We note that the errors given above correspond to a standard deviation. In SM, we discuss a more rigorous statistical error analysis of finite counting statistics. There, employing entanglement of GME-dimension two as the null hypothesis, we show that our method can observe GME with high confidence while using very few counts, e.g. in the case of 4-qutrit, a number of total counts $N \simeq 300$ will be already enough to get a $p$-value of order $10^{-5}$.

\subsection*{Non-locality}

\begin{figure}[tbp!]
\includegraphics [width=.48\textwidth]{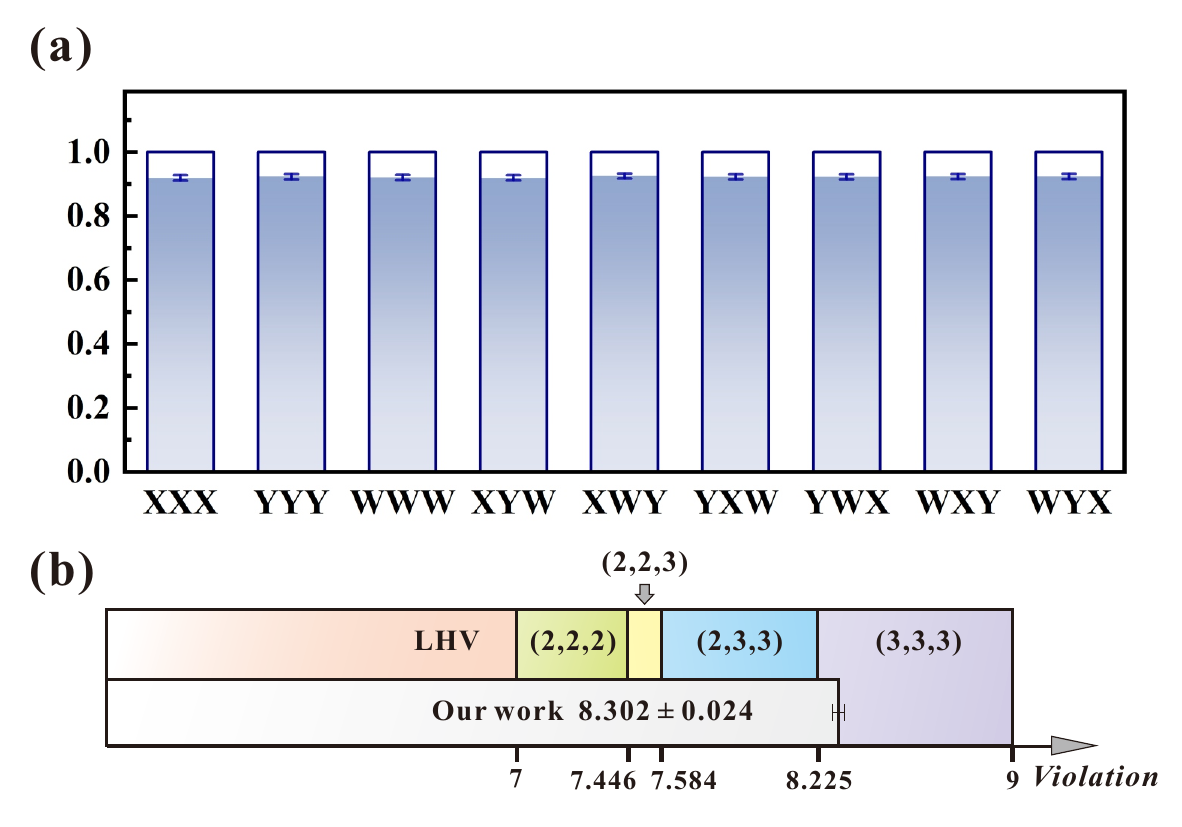}
\vspace{-0.3cm}%
\caption{\textbf{Experimental results for high-dimensional multi-partite Bell inequality.} (a) Experimental results under different measurement bases. Here, $X,Y,Z$ stands for three different project measurements on a single qutrit, and there are a total 9 sets of measurements for the Bell inequality. (b) Violation of the Bell inequality. The different colors represent different levels of violation of inequality. Our result violates the bounds of (2,3,3) and proves the existence of genuine high-dimensional multi-partite non-locality.}
\label{fig:plot_mermin}
\vspace{0.2cm}%
\end{figure}

Going beyond high-dimensional multi-partite entanglement detection,
%In addition to proving that the GHZ we prepared \MH{exhibits correlations beyond what is possible for any system of qubits}, more importantly, 
we now consider the possibility of experimentally observing high-dimensional multi-partite Bell correlations. 
%also want to see how the experimentally generated GHZ quantum states will perform in genuine high-dimensional multi-body nonlocality testing. 
Specifically, we set out to experimentally violate the chain of Bell-type inequalities given in Eq.~\eqref{nonlocalitybounds} using three-qutrit GHZ states. %We measured the Bell operator for three photons in three dimensions. 
The Bell operator requires us to measure nine global settings, each corresponding to a particular choice of input for the three photons, which we name $\{X,Y,W\}$ respectively (details see Section 6 in the SM). We obtain $\mathcal{B}=8.302 \pm 0.024$, which is 54.3 standard deviations above the LHV bounds, 7 standard deviations above the (2,2,3)-bound and 3.2 standard deviations above the (2,3,3) bound, see \eqref{nonlocalitybounds}. In this case, the estimation of the $p$-values requires more counts than the GME witness to observe significant violations. Therefore, we recorded a total of 10,143 counts to achieve a $p\text{-value} < 10^{-2}$ for the violation of the $(2,3,3)$ inequality. The experimental data and violations are displayed in Figure~\ref{fig:plot_mermin}. These results represent our successful foray into the observation of genuine high-dimensional multi-partite quantum non-locality. 
%In SM, we also compute the $p$-values of our various violations. The Bell inequality requires more counts than the GME witness to observe significant violations. Here we recorded a total of 10,143 counts and achieved a $p$-value of $8 \times 10^{-3}$ for the violation of  the (2,3,3) inequality.

%\section*{Discussion}

\section*{Discussion}
Our experiment demonstrates the multi-partite Bell correlations strong enough to falsify the constraints of qubit-based quantum theory in non-locality experiments. This was possible thanks to the high GHZ-state fidelities achievable in our path-identity scheme. This experiment marks an initial advancement towards realizing device-independent quantum information processing in the high-dimensional and multi-partite system. For that purpose, future challenges will involve demonstrations of such non-locality while also closing the detection and/or locality loopholes, while also performing multi-outcome measurements. A natural further step is to consider nonlocality with more than three parties in high dimensions. While in the four-party case our setup achieves potentially sufficient fidelity, the test requires development of inequalities analogous to \eqref{nonlocalitybounds} that exhibit also significant noise-tolerance. Even for standard LHV models, little is known about such inequalities.

%A related challenge is that of identifying more noise-tolerant Bell inequalities for many parties in high dimensions.

The visibility of HOM interference observed in the experiment is much higher than the fidelity of the state we successfully prepared. In principle, the quantum state we prepared can achieve higher fidelity. In the future, the use of better compensation techniques and stable structures will greatly improve the fidelity of our quantum state preparation, ensuring that the quantum state we prepared can be better utilised for various quantum information tasks and fundamental quantum physics problems \supercite{zhong2020quantum}. Our experimental platform has excellent scalability, which comes from the two-dimensional expansion of our photon pairs on a plane, and compared to one-dimensional expansion schemes, this scheme has higher path exchange efficiency \supercite{PhysRevLett.125.090503}. At the same time, the excellent indistinguishability of our light source and the high isolation of path switching ensure high-quality state preparation fidelity. 

Most entanglements in the real world are in high-dimensional entanglement. 
%The high-fidelity high-dimensional multi-photon entangled states we prepare will bring us new understanding of quantum physics \supercite{erhard2020advances}, such as new and more complex non-locality models and quantum teleportation of all quantum information of particles. 
The high-fidelity, high-dimensional multi-photon entangled states we prepare facilitate deeper insights into quantum physics\supercite{erhard2020advances}, including the investigation of increasingly complex nonlocality models and the realization of quantum teleportation involving the complete information of individual particles.
Our experimental platform is very suitable for implementing path identity schemes \supercite{PhysRevLett.119.240403, PhysRevA.99.032338}. Combined with high-dimensional Bell measurements \supercite{PhysRevLett.125.230501,Xing:23}, we can achieve the preparation of high-dimensional multi-photon GHZ states and graph states. These states can be utilized for high-dimensional quantum communication \supercite{PhysRevA.59.1829}, quantum sensing \supercite{liu2021distributed}, and quantum computing \supercite{lanyon2009simplifying} with high noise resistance. 

\subsection*{Data availability} 
The data that support the findings of this study are available from the corresponding authors upon reasonable request. 

\subsection*{Code availability} 
The custom codes used to produce the results presented in this paper are available from the corresponding authors upon request.

\printbibliography 
\noindent 
\subsection*{Acknowledgements}
We thank Mateus Ara\'ujo for helpful discussions.
The group in USTC was supported by the Innovation Program for Quantum Science and Technology (No. 2024ZD0301400, 2021ZD0301200), the NSFC (No.~62322513, No.~12374338, No.~11904357, No.~12174367, No.~12204458),   Anhui Provincial Natural Science Foundation (No. 2408085JX002), Anhui Province Science and Technology Innovation Project (No. 202423r06050004), China Postdoctoral Science Foundation (2021M700138).
N.A.,~G.C.,~and A.T. are supported by the Wenner-Gren Foundation, by the Swedish Research Council under Contract No. 2023-03498 and the Knut and Alice Wallenberg Foundation through the Wallenberg Center for Quantum Technology (WACQT). X.G. is supported by the Natural Science Foundation of Jiangsu Province (No. BK20233001). This work was partially carried out at the USTC Center for Micro and Nanoscale Research and Fabrication. 

\subsection*{Author Contributions Statement}

X.-M.H. and C.-X.H. contributed equally to this work. B.-H.L., C.-F.L. and G.-C.G. proposed the framework of the project. B.-H.L., X.-M.H. and C.-X.H. designed and carried out the experiment with the help of C.Z., Y.G., and Y.-F.H.. A.T., N.A. and G.C. developed the new Bell inequality and performed the theoretical analysis with the help of X.G. and M.H.. X.-M.H., A.T. and B.-H.L. managed the project. All authors discussed the results and contributed to the manuscript.

\subsection*{Competing Interests Statement} The authors declare no competing interests.

\clearpage
\end{document}